\documentclass[twocolumn,amsmath,amssymb]{revtex4-2}
\usepackage{graphicx}
\usepackage{dcolumn}
\usepackage{bm}
\usepackage{graphicx}
\usepackage{hyperref}
\usepackage[left=2cm, right=2cm, top=3.5cm, bottom=2cm]{geometry}

\begin{document}

\title{{\Large Investigating the effects of acceptor removal mechanism and impact ionization on proton irradiated 300 $\mu$m thick LGAD}}

\author{\large Rajiv Gupta$^1$}
\author{\large Sunidhi Saxena$^1$}
\author{\large Chakresh Jain$^2$}
\author{\large Kalpna Tiwari$^2$}
\author{\large Rahul Sharma$^2$}
\author{\large Namrata Agrawal$^3$}
\author{\large Ashutosh Bhardwaj$^2$}
\author{\large Kirti Ranjan$^2$}
\author{\large Ajay Kumar$^1$}
\email{ajay.ehep@gmail.com; ajay.phy@bhu.ac.in}
\affiliation{$^1$Banaras Hindu University (BHU), Department of Physics, Varanasi, Uttar Pradesh, India }
\affiliation{$^2$CDRST, Department of Physics and Astrophysics, University of Delhi, India}
\affiliation{$^3$Swami Shraddhanand College, University of Delhi, India}
\maketitle
\section{Introduction}

The High Luminosity Large Hadron Collider (HL-LHC), starting in 2029, aims for 4000 $fb^{-1}$ luminosity over 10 years. During its operations, CMS's Minimum Ionizing Particle Timing Detector (MTD) \cite{CERN-LHCC-2017-027} and ATLAS's High Granularity Timing Detector (HGTD) \cite{CERN-LHCC-2018-023} will use Low Gain Avalanche Detector (LGAD) technology \cite{PELLEGRINI201412} for improved timing measurements and particle detection performance.

LGADs are semiconductor detectors that use a high electric field ($>200$ kV/cm) to achieve moderate charge multiplication. On a p-type substrate, a moderately doped p-type layer is implanted beneath a highly doped n-type layer. LGADs have enhanced charge collection compared to standard PIN diodes because of the additional p-type layer (also called Gain Layer), leading to stronger signals via impact ionization. However, they are sensitive to radiation damage, including that from proton irradiation \cite{Kramberger_2015}.

Proton irradiation degrades the gain layer \cite{rivera2023gainlayerdegradationstudy} by reducing the gain layer concentration with irradiation (also called  Acceptor Removal Mechanism (ARM)), and introduces trap defects in LGADs, leading to reduced signal. The impact of proton irradiation on PIN diodes can be modeled using Technology Computer Aided Design (TCAD) Silvaco \cite{ATLAS}, with one such model being the Proton Damage Model (PDM), developed by the University of Delhi \cite{inproceedings} which can estimate leakage current, full depletion voltage and charge collection efficiency.

This study extends the validation of the PDM, previously applied to PIN diodes \cite{inproceedings}, to LGADs. Since the gain layer gets affected by the irradiation and the charge multiplication process of LGAD is sensitive to the impact ionization coefficients, the present work aims to include these two effects to the PDM. The model of gain layer degradation is adopted from Ref.\cite{rivera2023gainlayerdegradationstudy}, and the impact ionization coefficients are optimized in a manner similar to the work reported in Ref.\cite{impact}. By accounting for trap defects, gain layer degradation, and optimized impact ionization coefficients, the charge collection predictions for irradiated LGADs are significantly improved in modeling the proton damage in LGAD. The experimental results of charge collection for both irradiated and non-irradiated LGADs have been obtained from Ref.\cite{Kramberger_2015}. \\

\begin{figure}
\includegraphics[width=55mm]{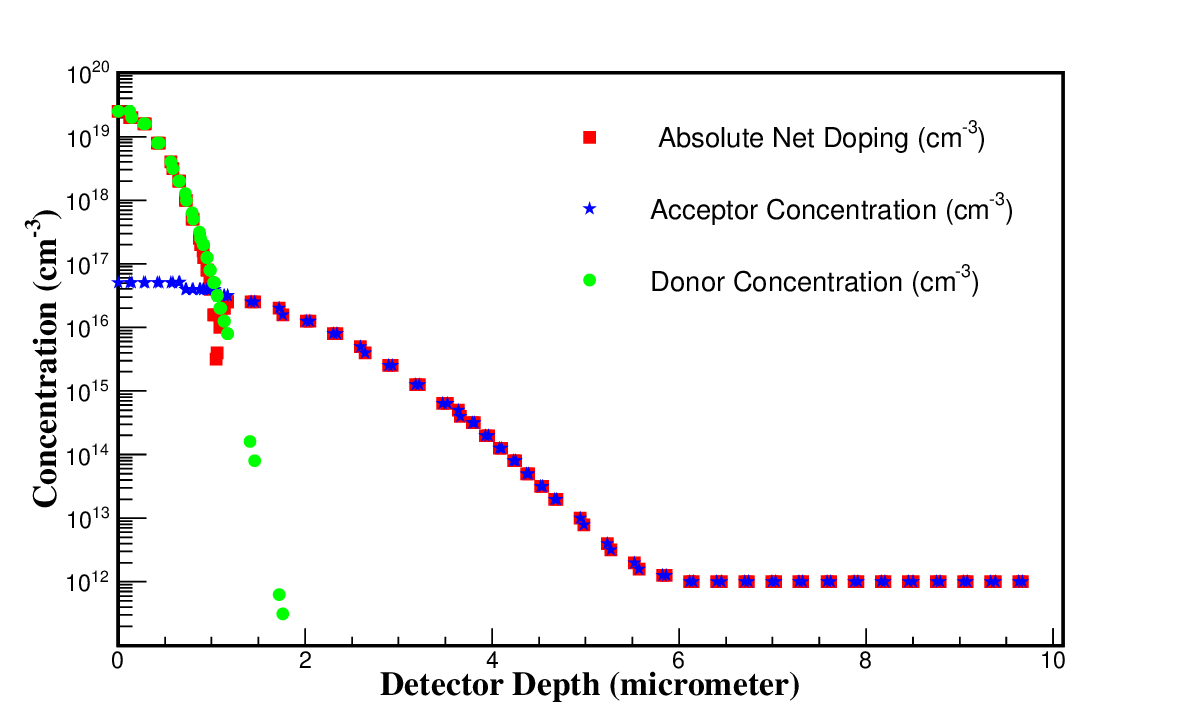}
\caption{\label{fig1} Doping Profile of simulated LGAD.}
\end{figure}

\section{Non-Irradiated LGAD ($\phi=0$)}

The LGAD designed in Silvaco features a 2D plane-parallel geometry with $n^{++}$ / $p^{+}$ / p / $p^{++}$ configuration. The area of the sensor simulated is $5mm \times 5mm$ with a strip width of 80 $\mu$m and thickness of 300 $\mu$m. The bulk of the sensor is p-type with uniform doping of $1 \times 10^{12}$ $cm^{-3}$. The doping profile of the LGAD structure used in simulation is shown in Fig. \ref{fig1}.

\begin{figure}
\includegraphics[width=55mm]{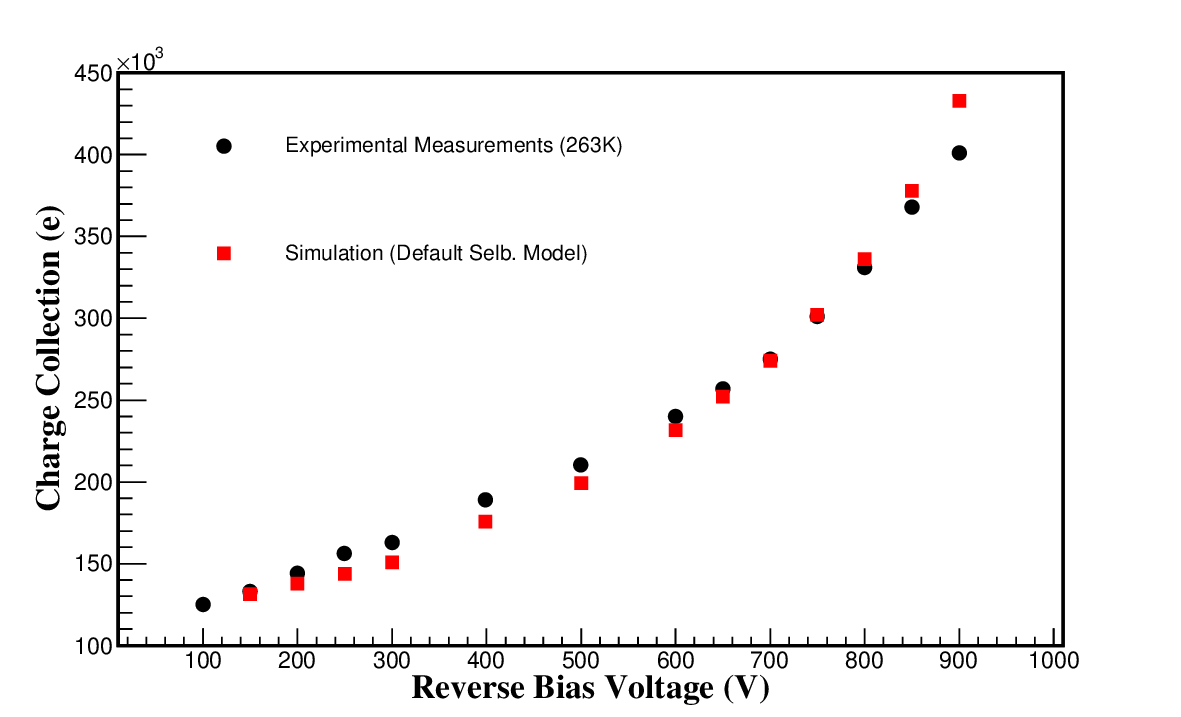}
\caption{\label{fig2} Charge Collections as a function of bias voltage for non-irradiated LGAD. Experimental measurements are obtained from the reference \cite{Kramberger_2015}.}
\end{figure}

A 660 nm red laser is shone from the backside ($p^{++}$ end) for charge collection. The simulated charge collection for a non-irradiated LGAD, using the default Selberherr impact ionization model, closely matches the experimental data \cite{Kramberger_2015}, validating the doping profile and other parameters used in this work (see Fig. \ref{fig2}) for the non-irradiated LGAD.

\section{Simulation of Proton Irradiated LGAD ($\phi$ = $4.9 \times 10^{14}$ 1 $MeV$ $n_{eq}$ $cm^{-2}$)}

For ARM, it is considered that the acceptors in the gain layer decrease exponentially with an increase in proton fluence and follow a single exponential distribution.
\begin{center}
 $N_{P}(\phi) = Np(0)exp(-C_{p}\phi)$ \hspace{0.5cm} (1)  
\end{center}
where, $N_{P}(\phi)$ is the gain layer concentration at a certain incident 1 MeV neutron equivalent proton fluence ($\phi$), $Np(0)$ is the initial acceptor concentration in the gain layer and $C_{p}$ is the acceptor removal constant for proton irradiation.

The doping concentration value of the gain layer extracted from Eqn.1 is incorporated in the simulation for fluence, $\phi$ = $4.9 \times 10^{14}$ 1 $MeV$ $n_{eq}$ $cm^{-2}$. The trap defects are introduced by incorporating the two trap PDM \cite{inproceedings} into the modeling. The value of $C_{p}$ to incorporate ARM is $16 \times 10^{-16}$ $ cm^{2}$ which is the same value mentioned for the experimental measurements \cite{Kramberger_2015}.

The Selberherr's impact ionization rates for electrons and holes are modeled in Silvaco using Eqn.2, which is based on the classical Chynoweth model \cite{PhysRev.109.1537}. The impact ionization coefficients $\alpha(n,p)$ are given by:
\begin{equation*}
\alpha(n,p) = A_{(n,p)}exp[-\frac{B_{(n,p)}}{E}]  \hspace{0.5cm} (2)
\end{equation*}
where, E is the electric field, $A_{(n,p)}$ and $B_{(n,p)}$ are the four important model parameters for impact ionization of electrons (n) and holes (p). The default values of the parameters used in simulations are taken from Ref.\cite{ATLAS}.

\begin{figure}
\includegraphics[width=75mm]{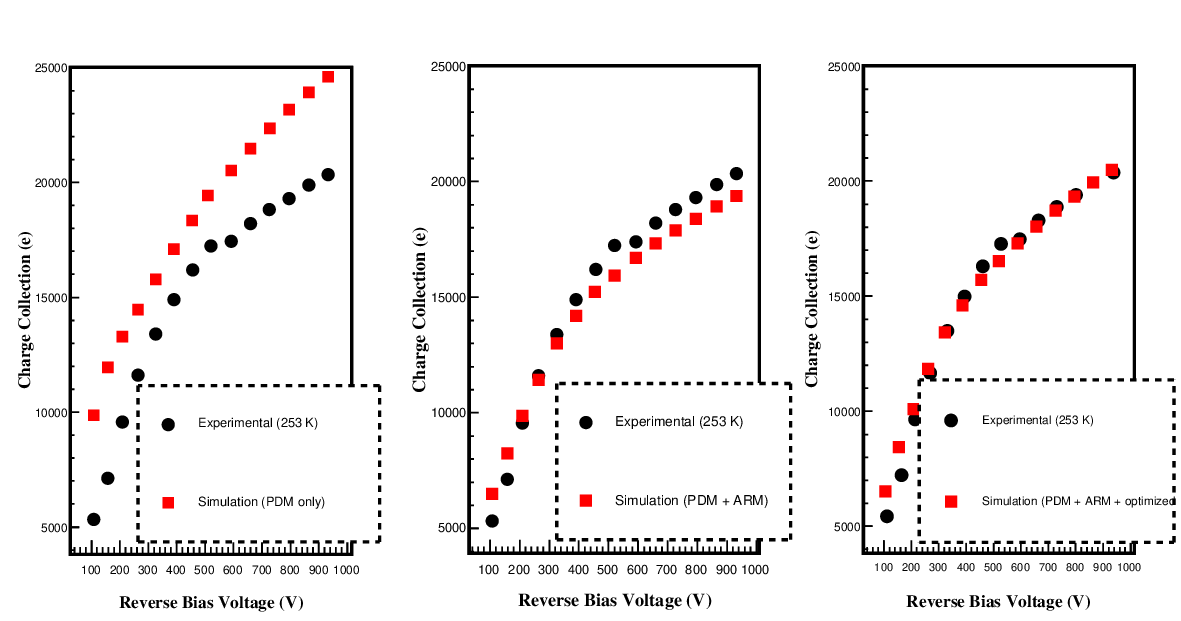}
\caption{\label{fig3} Comparison between the simulated charge collection and measured charge collection \cite{Kramberger_2015} for a proton irradiated LGAD with a fluence $4.9 \times 10^{14}$ 1 MeV $n_{eq}$ $cm^{-2}$. Simulation incorporating only PDM (left), simulation incorporating PDM and ARM (centre), and simulation incorporating PDM, ARM, and optimized Selb. parameter (right).}
\end{figure}

When simulating the designed LGAD with only PDM, the charge collection is overestimated Fig. \ref{fig3} (left). Incorporating ARM along with PDM improves the agreement; however, it leads to a slight underestimation of the charge collection at high voltages Fig. \ref{fig3} (centre). The parameter sensitivity of the Selberherr impact ionization model showed that optimizing the coefficient $B_{p}$ from $2.036 \times 10^{6}$ V/cm to $1.30 \times 10^{6}$ V/cm alone improves charge collection agreement with measurement Fig. \ref{fig3} (right). The charge collections obtained from optimized parameters for irradiated LGAD are shown in Fig. \ref{fig3} (right). This results in the overall good agreement between modeling and measurements.


\section*{Summary}
This work enhances the proton damage model developed for PIN diodes to the LGADs. It is observed that by incorporating ARM and optimizing impact ionization coefficients, a better agreement between modeling and measurement results is obtained.
\section*{Acknowledgments}
The authors sincerely acknowledge financial support from the Institute of Eminence (IoE), BHU Grant number 6031, and the DST. RG, SS, and KT acknowledge UGC for financial support.





\end{document}